\documentclass[aps,prl,twocolumn]{revtex4}  \usepackage{graphicx}
\usepackage{amsmath}

\def\PsfigVersion{1.9}
\ifx\undefined\psfig\else \fi

\let\LaTeXAtSign=\@
\let\@=\relax
\edef\psfigRestoreAt{\catcode`\@=\number\catcode`@\relax}
\catcode`\@=11\relax
\newwrite\@unused
\def\ps@typeout#1{{\let\protect\string\immediate\write\@unused{#1}}}
\ps@typeout{psfig/tex \PsfigVersion}

\def\figurepath{./}

\def\@nnil{\@nil}
\def\@empty{}
\def\@psdonoop#1\@@#2#3{}
\def\@psdo#1:=#2\do#3{\edef\@psdotmp{#2}\ifx\@psdotmp\@empty \else
    \expandafter\@psdoloop#2,\@nil,\@nil\@@#1{#3}\fi}
\def\@psdoloop#1,#2,#3\@@#4#5{\def#4{#1}\ifx #4\@nnil \else
       #5\def#4{#2}\ifx #4\@nnil \else#5\@ipsdoloop #3\@@#4{#5}\fi\fi}
\def\@ipsdoloop#1,#2\@@#3#4{\def#3{#1}\ifx #3\@nnil 
       \let\@nextwhile=\@psdonoop \else
      #4\relax\let\@nextwhile=\@ipsdoloop\fi\@nextwhile#2\@@#3{#4}}
\def\@tpsdo#1:=#2\do#3{\xdef\@psdotmp{#2}\ifx\@psdotmp\@empty \else
    \@tpsdoloop#2\@nil\@nil\@@#1{#3}\fi}
\def\@tpsdoloop#1#2\@@#3#4{\def#3{#1}\ifx #3\@nnil 
       \let\@nextwhile=\@psdonoop \else
      #4\relax\let\@nextwhile=\@tpsdoloop\fi\@nextwhile#2\@@#3{#4}}
\ifx\undefined\fbox
\newdimen\fboxrule
\newdimen\fboxsep
\newdimen\ps@tempdima
\newbox\ps@tempboxa
\long\def\fbox#1{\leavevmode\setbox\ps@tempboxa\hbox{#1}\ps@tempdima\fboxrule
    \advance\ps@tempdima \fboxsep \advance\ps@tempdima \dp\ps@tempboxa
   \hbox{\lower \ps@tempdima\hbox
  {\vbox{\hrule height \fboxrule
          \hbox{\vrule width \fboxrule \hskip\fboxsep
          \vbox{\vskip\fboxsep \box\ps@tempboxa\vskip\fboxsep}\hskip 
                 \fboxsep\vrule width \fboxrule}
                 \hrule height \fboxrule}}}}
\fi
\newread\ps@stream
\newif\ifnot@eof       
\newif\if@noisy        
\newif\if@atend        
\newif\if@psfile       
{\catcode`\%=12\global\gdef\epsf@start{
\def\epsf@PS{PS}
\def\epsf@getbb#1{%
\openin\ps@stream=#1
\ifeof\ps@stream\ps@typeout{Error, File #1 not found}\else
   {\not@eoftrue \chardef\other=12
    \def\do##1{\catcode`##1=\other}\dospecials \catcode`\ =10
    \loop
       \if@psfile
	  \read\ps@stream to \epsf@fileline
       \else{
	  \obeyspaces
          \read\ps@stream to \epsf@tmp\global\let\epsf@fileline\epsf@tmp}
       \fi
       \ifeof\ps@stream\not@eoffalse\else
       \if@psfile\else
       \expandafter\epsf@test\epsf@fileline:. \\%
       \fi
          \expandafter\epsf@aux\epsf@fileline:. \\%
       \fi
   \ifnot@eof\repeat
   }\closein\ps@stream\fi}%
\long\def\epsf@test#1#2#3:#4\\{\def\epsf@testit{#1#2}
			\ifx\epsf@testit\epsf@start\else
\ps@typeout{Warning! File does not start with `\epsf@start'.  It may not be a PostScript file.}
			\fi
			\@psfiletrue} 
{\catcode`\%=12\global\let\epsf@percent=
\long\def\epsf@aux#1#2:#3\\{\ifx#1\epsf@percent
   \def\epsf@testit{#2}\ifx\epsf@testit\epsf@bblit
	\@atendfalse
        \epsf@atend #3 . \\%
	\if@atend	
	   \if@verbose{
		\ps@typeout{psfig: found `(atend)'; continuing search}
	   }\fi
        \else
        \epsf@grab #3 . . . \\%
        \not@eoffalse
        \global\no@bbfalse
        \fi
   \fi\fi}%
\def\epsf@grab #1 #2 #3 #4 #5\\{%
   \global\def\epsf@llx{#1}\ifx\epsf@llx\empty
      \epsf@grab #2 #3 #4 #5 .\\\else
   \global\def\epsf@lly{#2}%
   \global\def\epsf@urx{#3}\global\def\epsf@ury{#4}\fi}%
\def\epsf@atendlit{(atend)} 
\def\epsf@atend #1 #2 #3\\{%
   \def\epsf@tmp{#1}\ifx\epsf@tmp\empty
      \epsf@atend #2 #3 .\\\else
   \ifx\epsf@tmp\epsf@atendlit\@atendtrue\fi\fi}

\chardef\psletter = 11 
\chardef\other = 12

\newif \ifdebug 
\newif\ifc@mpute 
\c@mputetrue 

\let\then = \relax
\def\r@dian{pt }
\let\r@dians = \r@dian
\let\dimensionless@nit = \r@dian
\let\dimensionless@nits = \dimensionless@nit
\def\internal@nit{sp }
\let\internal@nits = \internal@nit
\newif\ifstillc@nverging
\def \Mess@ge #1{\ifdebug \then \message {#1} \fi}

{ 
	\catcode `\@ = \psletter
	\gdef \nodimen {\expandafter \n@dimen \the \dimen}
	\gdef \term #1 #2 #3%
	       {\edef \t@ {\the #1}
		\edef \t@@ {\expandafter \n@dimen \the #2\r@dian}%
		\t@rm {\t@} {\t@@} {#3}%
	       }
	\gdef \t@rm #1 #2 #3%
	       {{%
		\count 0 = 0
		\dimen 0 = 1 \dimensionless@nit
		\dimen 2 = #2\relax
		\Mess@ge {Calculating term #1 of \nodimen 2}%
		\loop
		\ifnum	\count 0 < #1
		\then	\advance \count 0 by 1
			\Mess@ge {Iteration \the \count 0 \space}%
			\Multiply \dimen 0 by {\dimen 2}%
			\Mess@ge {After multiplication, term = \nodimen 0}%
			\Divide \dimen 0 by {\count 0}%
			\Mess@ge {After division, term = \nodimen 0}%
		\repeat
		\Mess@ge {Final value for term #1 of 
				\nodimen 2 \space is \nodimen 0}%
		\xdef \Term {#3 = \nodimen 0 \r@dians}%
		\aftergroup \Term
	       }}
	\catcode `\p = \other
	\catcode `\t = \other
	\gdef \n@dimen #1pt{#1} 
}

\def \Divide #1by #2{\divide #1 by #2} 

\def \Multiply #1by #2
       {{
	\count 0 = #1\relax
	\count 2 = #2\relax
	\count 4 = 65536
	\Mess@ge {Before scaling, count 0 = \the \count 0 \space and
			count 2 = \the \count 2}%
	\ifnum	\count 0 > 32767 
	\then	\divide \count 0 by 4
		\divide \count 4 by 4
	\else	\ifnum	\count 0 < -32767
		\then	\divide \count 0 by 4
			\divide \count 4 by 4
		\else
		\fi
	\fi
	\ifnum	\count 2 > 32767 
	\then	\divide \count 2 by 4
		\divide \count 4 by 4
	\else	\ifnum	\count 2 < -32767
		\then	\divide \count 2 by 4
			\divide \count 4 by 4
		\else
		\fi
	\fi
	\multiply \count 0 by \count 2
	\divide \count 0 by \count 4
	\xdef \product {#1 = \the \count 0 \internal@nits}%
	\aftergroup \product
       }}

\def\r@duce{\ifdim\dimen0 > 90\r@dian \then   
		\multiply\dimen0 by -1
		\advance\dimen0 by 180\r@dian
		\r@duce
	    \else \ifdim\dimen0 < -90\r@dian \then  
		\advance\dimen0 by 360\r@dian
		\r@duce
		\fi
	    \fi}

\def\Sine#1%
       {{%
	\dimen 0 = #1 \r@dian
	\r@duce
	\ifdim\dimen0 = -90\r@dian \then
	   \dimen4 = -1\r@dian
	   \c@mputefalse
	\fi
	\ifdim\dimen0 = 90\r@dian \then
	   \dimen4 = 1\r@dian
	   \c@mputefalse
	\fi
	\ifdim\dimen0 = 0\r@dian \then
	   \dimen4 = 0\r@dian
	   \c@mputefalse
	\fi
	\ifc@mpute \then
		\divide\dimen0 by 180
		\dimen0=3.141592654\dimen0
		\dimen 2 = 3.1415926535897963\r@dian 
		\divide\dimen 2 by 2 
		\Mess@ge {Sin: calculating Sin of \nodimen 0}%
		\count 0 = 1 
		\dimen 2 = 1 \r@dian 
		\dimen 4 = 0 \r@dian 
		\loop
			\ifnum	\dimen 2 = 0 
			\then	\stillc@nvergingfalse 
			\else	\stillc@nvergingtrue
			\fi
			\ifstillc@nverging 
			\then	\term {\count 0} {\dimen 0} {\dimen 2}%
				\advance \count 0 by 2
				\count 2 = \count 0
				\divide \count 2 by 2
				\ifodd	\count 2 
				\then	\advance \dimen 4 by \dimen 2
				\else	\advance \dimen 4 by -\dimen 2
				\fi
		\repeat
	\fi		
			\xdef \sine {\nodimen 4}%
       }}

\def\Cosine#1{\ifx\sine\UnDefined\edef\Savesine{\relax}\else
		             \edef\Savesine{\sine}\fi
	{\dimen0=#1\r@dian\advance\dimen0 by 90\r@dian
	 \Sine{\nodimen 0}
	 \xdef\cosine{\sine}
	 \xdef\sine{\Savesine}}}	      

\def\psdraft{
	\def\@psdraft{0}
}
\def\psfull{
	\def\@psdraft{100}
}

\psfull

\newif\if@scalefirst
\def\psscalefirst{\@scalefirsttrue}
\def\psrotatefirst{\@scalefirstfalse}
\psrotatefirst

\newif\if@draftbox
\def\psnodraftbox{
	\@draftboxfalse
}
\def\psdraftbox{
	\@draftboxtrue
}
\@draftboxtrue

\newif\if@prologfile
\newif\if@postlogfile
\def\pssilent{
	\@noisyfalse
}
\def\psnoisy{
	\@noisytrue
}
\psnoisy
\newif\if@bbllx
\newif\if@bblly
\newif\if@bburx
\newif\if@bbury
\newif\if@height
\newif\if@width
\newif\if@rheight
\newif\if@rwidth
\newif\if@angle
\newif\if@clip
\newif\if@verbose
\newif\if@scale
\def\@p@@sclip#1{\@cliptrue}

\newif\if@decmpr

\def\@p@@sfigure#1{\def\@p@sfile{null}\def\@p@sbbfile{null}
	        \openin1=#1.bb
		\ifeof1\closein1
	        	\openin1=\figurepath#1.bb
			\ifeof1\closein1
			        \openin1=#1
				\ifeof1\closein1%
				       \openin1=\figurepath#1
					\ifeof1
					   \ps@typeout{Error, File #1 not found}
						\if@bbllx\if@bblly
				   		\if@bburx\if@bbury
			      				\def\@p@sfile{#1}%
			      				\def\@p@sbbfile{#1}%
							\@decmprfalse
				  	   	\fi\fi\fi\fi
					\else\closein1
				    		\def\@p@sfile{\figurepath#1}%
				    		\def\@p@sbbfile{\figurepath#1}%
						\@decmprfalse
	                       		\fi%
			 	\else\closein1%
					\def\@p@sfile{#1}
					\def\@p@sbbfile{#1}
					\@decmprfalse
			 	\fi
			\else
				\def\@p@sfile{\figurepath#1}
				\def\@p@sbbfile{\figurepath#1.bb}
				\@decmprtrue
			\fi
		\else
			\def\@p@sfile{#1}
			\def\@p@sbbfile{#1.bb}
			\@decmprtrue
		\fi}

\def\@p@@sfile#1{\@p@@sfigure{#1}}

\def\@p@@sbbllx#1{
		\@bbllxtrue
		\dimen100=#1
		\edef\@p@sbbllx{\number\dimen100}
}
\def\@p@@sbblly#1{
		\@bbllytrue
		\dimen100=#1
		\edef\@p@sbblly{\number\dimen100}
}
\def\@p@@sbburx#1{
		\@bburxtrue
		\dimen100=#1
		\edef\@p@sbburx{\number\dimen100}
}
\def\@p@@sbbury#1{
		\@bburytrue
		\dimen100=#1
		\edef\@p@sbbury{\number\dimen100}
}
\def\@p@@sheight#1{
		\@heighttrue
		\dimen100=#1
   		\edef\@p@sheight{\number\dimen100}
}
\def\@p@@swidth#1{
		\@widthtrue
		\dimen100=#1
		\edef\@p@swidth{\number\dimen100}
}
\def\@p@@srheight#1{
		\@rheighttrue
		\dimen100=#1
		\edef\@p@srheight{\number\dimen100}
}
\def\@p@@srwidth#1{
		\@rwidthtrue
		\dimen100=#1
		\edef\@p@srwidth{\number\dimen100}
}
\def\@p@@sangle#1{
		\@angletrue
		\edef\@p@sangle{#1} 
}
\def\@p@@srotate#1{\@p@@sangle{-#1}}
\def\@p@@sscale#1{
		\@scaletrue
		\edef\@p@sscale{#1}
}
\def\@p@@ssilent#1{ 
		\@verbosefalse
}
\def\@p@@sprolog#1{\@prologfiletrue\def\@prologfileval{#1}}
\def\@p@@spostlog#1{\@postlogfiletrue\def\@postlogfileval{#1}}
\def\@cs@name#1{\csname #1\endcsname}
\def\@setparms#1=#2,{\@cs@name{@p@@s#1}{#2}}
\def\ps@init@parms{
		\@bbllxfalse \@bbllyfalse
		\@bburxfalse \@bburyfalse
		\@heightfalse \@widthfalse
		\@rheightfalse \@rwidthfalse
		\@scalefalse
		\def\@p@sbbllx{}\def\@p@sbblly{}
		\def\@p@sbburx{}\def\@p@sbbury{}
		\def\@p@sheight{}\def\@p@swidth{}
		\def\@p@srheight{}\def\@p@srwidth{}
		\def\@p@sangle{0}
		\def\@p@sfile{} \def\@p@sbbfile{}
		\def\@p@scost{10}
		\def\@sc{}
		\@prologfilefalse
		\@postlogfilefalse
		\@clipfalse
		\if@noisy
			\@verbosetrue
		\else
			\@verbosefalse
		\fi
}
\def\parse@ps@parms#1{
	 	\@psdo\@psfiga:=#1\do
		   {\expandafter\@setparms\@psfiga,}}
\newif\ifno@bb
\def\bb@missing{
	\if@verbose{
		\ps@typeout{psfig: searching \@p@sbbfile \space  for bounding box}
	}\fi
	\no@bbtrue
	\epsf@getbb{\@p@sbbfile}
        \ifno@bb \else \bb@cull\epsf@llx\epsf@lly\epsf@urx\epsf@ury\fi
}	
\def\bb@cull#1#2#3#4{
	\dimen100=#1 bp\edef\@p@sbbllx{\number\dimen100}
	\dimen100=#2 bp\edef\@p@sbblly{\number\dimen100}
	\dimen100=#3 bp\edef\@p@sbburx{\number\dimen100}
	\dimen100=#4 bp\edef\@p@sbbury{\number\dimen100}
	\no@bbfalse
}
\newdimen\p@intvaluex
\newdimen\p@intvaluey
\def\rotate@#1#2{{\dimen0=#1 sp\dimen1=#2 sp
		  \global\p@intvaluex=\cosine\dimen0
		  \dimen3=\sine\dimen1
		  \global\advance\p@intvaluex by -\dimen3
		  \global\p@intvaluey=\sine\dimen0
		  \dimen3=\cosine\dimen1
		  \global\advance\p@intvaluey by \dimen3
		  }}
\def\compute@bb{
		\no@bbfalse
		\if@bbllx \else \no@bbtrue \fi
		\if@bblly \else \no@bbtrue \fi
		\if@bburx \else \no@bbtrue \fi
		\if@bbury \else \no@bbtrue \fi
		\ifno@bb \bb@missing \fi
		\ifno@bb \ps@typeout{FATAL ERROR: no bb supplied or found}
			\no-bb-error
		\fi
		\count203=\@p@sbburx
		\count204=\@p@sbbury
		\advance\count203 by -\@p@sbbllx
		\advance\count204 by -\@p@sbblly
		\edef\ps@bbw{\number\count203}
		\edef\ps@bbh{\number\count204}
		\if@angle 
			\Sine{\@p@sangle}\Cosine{\@p@sangle}
	        	{\dimen100=\maxdimen\xdef\r@p@sbbllx{\number\dimen100}
					    \xdef\r@p@sbblly{\number\dimen100}
			                    \xdef\r@p@sbburx{-\number\dimen100}
					    \xdef\r@p@sbbury{-\number\dimen100}}
                        \def\minmaxtest{
			   \ifnum\number\p@intvaluex<\r@p@sbbllx
			      \xdef\r@p@sbbllx{\number\p@intvaluex}\fi
			   \ifnum\number\p@intvaluex>\r@p@sbburx
			      \xdef\r@p@sbburx{\number\p@intvaluex}\fi
			   \ifnum\number\p@intvaluey<\r@p@sbblly
			      \xdef\r@p@sbblly{\number\p@intvaluey}\fi
			   \ifnum\number\p@intvaluey>\r@p@sbbury
			      \xdef\r@p@sbbury{\number\p@intvaluey}\fi
			   }
			\rotate@{\@p@sbbllx}{\@p@sbblly}
			\minmaxtest
			\rotate@{\@p@sbbllx}{\@p@sbbury}
			\minmaxtest
			\rotate@{\@p@sbburx}{\@p@sbblly}
			\minmaxtest
			\rotate@{\@p@sbburx}{\@p@sbbury}
			\minmaxtest
			\edef\@p@sbbllx{\r@p@sbbllx}\edef\@p@sbblly{\r@p@sbblly}
			\edef\@p@sbburx{\r@p@sbburx}\edef\@p@sbbury{\r@p@sbbury}
		\fi
		\count203=\@p@sbburx
		\count204=\@p@sbbury
		\advance\count203 by -\@p@sbbllx
		\advance\count204 by -\@p@sbblly
		\edef\@bbw{\number\count203}
		\edef\@bbh{\number\count204}
}
\def\in@hundreds#1#2#3{\count240=#2 \count241=#3
		     \count100=\count240	
		     \divide\count100 by \count241
		     \count101=\count100
		     \multiply\count101 by \count241
		     \advance\count240 by -\count101
		     \multiply\count240 by 10
		     \count101=\count240	
		     \divide\count101 by \count241
		     \count102=\count101
		     \multiply\count102 by \count241
		     \advance\count240 by -\count102
		     \multiply\count240 by 10
		     \count102=\count240	
		     \divide\count102 by \count241
		     \count200=#1\count205=0
		     \count201=\count200
			\multiply\count201 by \count100
		 	\advance\count205 by \count201
		     \count201=\count200
			\divide\count201 by 10
			\multiply\count201 by \count101
			\advance\count205 by \count201
		     \count201=\count200
			\divide\count201 by 100
			\multiply\count201 by \count102
			\advance\count205 by \count201
		     \edef\@result{\number\count205}
}
\def\ps@scaleinhundreds#1{
		\in@hundreds{#1}{\@p@sscale}{100}
		\edef#1{\@result}
}
\def\compute@wfromh{
		\in@hundreds{\@p@sheight}{\@bbw}{\@bbh}
		\edef\@p@swidth{\@result}
}
\def\compute@hfromw{
	        \in@hundreds{\@p@swidth}{\@bbh}{\@bbw}
		\edef\@p@sheight{\@result}
}
\def\compute@handw{
		\if@height 
			\if@width
			\else
				\compute@wfromh
			\fi
		\else 
			\if@width
				\compute@hfromw
			\else
				\edef\@p@sheight{\@bbh}
				\edef\@p@swidth{\@bbw}
			\fi
		\fi
}
\def\compute@resv{
		\if@rheight \else \edef\@p@srheight{\@p@sheight} \fi
		\if@rwidth \else \edef\@p@srwidth{\@p@swidth} \fi
}
\def\compute@sizes{
	\compute@bb
	\if@scalefirst\if@angle
	\if@width
	   \in@hundreds{\@p@swidth}{\@bbw}{\ps@bbw}
	   \edef\@p@swidth{\@result}
	\fi
	\if@height
	   \in@hundreds{\@p@sheight}{\@bbh}{\ps@bbh}
	   \edef\@p@sheight{\@result}
	\fi
	\fi\fi
	\compute@handw
	\compute@resv
	\if@scale
	   \if@verbose
	      \ps@typeout{(scaling by \@p@sscale)}%
	   \fi
	   \ps@scaleinhundreds{\@p@swidth}%
	   \ps@scaleinhundreds{\@p@sheight}%
	   \ps@scaleinhundreds{\@p@srwidth}%
	   \ps@scaleinhundreds{\@p@srheight}%
	\fi
}

\def\psfig#1{\vbox {
	\ps@init@parms
	\parse@ps@parms{#1}
	\compute@sizes
	\ifnum\@p@scost<\@psdraft{
		\special{ps::[begin] 	\@p@swidth \space \@p@sheight \space
				\@p@sbbllx \space \@p@sbblly \space
				\@p@sbburx \space \@p@sbbury \space
				startTexFig \space }
		\if@angle
			\special {ps:: \@p@sangle \space rotate \space} 
		\fi
		\if@clip{
			\if@verbose{
				\ps@typeout{(clip)}
			}\fi
			\special{ps:: doclip \space }
		}\fi
		\if@prologfile
		    \special{ps: plotfile \@prologfileval \space } \fi
		\if@decmpr{
			\if@verbose{
				\ps@typeout{psfig: including \@p@sfile.Z \space }
			}\fi
			\special{ps: plotfile "`zcat \@p@sfile.Z" \space }
		}\else{
			\if@verbose{
				\ps@typeout{psfig: including \@p@sfile \space }
			}\fi
			\special{ps: plotfile \@p@sfile \space }
		}\fi
		\if@postlogfile
		    \special{ps: plotfile \@postlogfileval \space } \fi
		\special{ps::[end] endTexFig \space }
		\vbox to \@p@srheight true sp{
			\hbox to \@p@srwidth true sp{
				\hss
			}
		\vss
		}
	}\else{
		\if@draftbox{		
			\hbox{\frame{\vbox to \@p@srheight true sp{
			\vss
			\hbox to \@p@srwidth true sp{ \hss \@p@sfile \hss }
			\vss
			}}}
		}\else{
			\vbox to \@p@srheight true sp{
			\vss
			\hbox to \@p@srwidth true sp{\hss}
			\vss
			}
		}\fi

	}\fi
}}
\psfigRestoreAt
\let\@=\LaTeXAtSign

\begin{document}

\title{Conserved geometric phase and group velocity}   \author{S. Selenu}

\affiliation{Atomistic Simulation Centre, School of Mathematics and
 Physics\\ Queen's University Belfast, Belfast BT7 1NN, Northern
 Ireland, UK}

\begin{abstract}
In this paper we make use of the concept of conserved geometric phase and of group velocity,in conjunction with the representation 
theory\cite{Dirac,Dirac-letture}, in order to derive some relevant physical quantities for the description of the dielectric and magnetic 
response of crystalline materials.  
As an application of the model, we derive the expression of the macroscopic dipole moment per unit volume, and the expression of the 
current induced by a uniform static external electromagnetic field. 

\end{abstract}

\date{\today} \maketitle

\section{Introduction}
\label{AAphase} 

In 1983, in a milestone paper, M.V. Berry\cite{Berry} showed that a quantal
system in an eigenstate, slowly transported round a circuit $C$, by varying
parameters $\beta$ in its Hamiltonian $\tilde{H}_{\beta}$, will aquire a
geometrical phase factor $\Phi$ in addition to the dynamical phase
factor. 

Moreover, in 1986 Y. Aharonov and J. Anandan\cite{Aronov} showed that it is possible 
to define a new geometrical phase factor for $\it{any}$ cyclic evolution of
a quantum system, independently of the adiabatic approximation\cite{Berry} and
expressible as a gauge invariant quantity, including Berry's case
as a particular case.  

In 1988, J.Zak\cite{ZakBerry} showed also that it is possible to define a
Berry's phase for the dynamics of electrons in infinite periodic systems,
suggesting how to label bands of a crystal by making use of the conserved
geometric phase.

In early nineties R.Resta and D.Vanderbilt \cite{KS-V,Resta2} showed,  
how to relate the conserved geometric phase to the macroscopic dipole moment per unit volume of a crystalline insulator. 
The model have been called by the authors the modern theory of polarization.  
In the following section, we shall review the concept of polarization in crystals, within the modern point of view\cite{Resta2,Resta1}. 

\section{Polarization and geometric phase}
\label{PolasBerry}
\noindent The problem of the $\it{dielectric}$ and $\it{magnetic}$ response of matter, in condensed matter
physics, have been intensively studied in the last
decades\cite{Nenciu}-\cite{Martin2} (and references therein).
A particularly interesting example is the treatment of the electronic
structure of a crystalline material in an external uniform electrostatic field.
The modern theory of polarization in crystals\cite{Resta1,Resta2} is one of the most
important recent developments, in fact,  
the macroscopic polarization is a basic quantity used to describe, classically, dielectric
media. 

However a crystal is only a $\it{particular}$ example of a quantal body and the $\it{classical}$ concept of polarization extends also to
non crystalline (disordered) systems, and it should not depend on the particular model system employed
to perform explicit calculations. 

As shown by Resta\cite{Resta-ferro} it is possible to define variations $\Delta P$, of the dipole per unit
volume of a crystal, due to a $\it{source}$, and given by 

\begin{equation}
\label{equation1}
\Delta P = P^{(1)}-P^{(0)}=\int^{1}_{0} ( \frac{\partial P}{\partial
\lambda})d\lambda
\end{equation}

\noindent Here $\lambda$ is a dimensionless parameter depending on time;
$\lambda=0$ and $\lambda=1$ represent the initial and the final states of the
system, along a transformation that changes the periodic potential.
 
An explicit representation of $P^{(\lambda)}$ in terms of the
amplitudes $u_{n\bf{k}}$ of Bloch-states\cite{Ashcroft,Kittel,Martin-book}, has been derived by King-Smith and
Vanderbilt \cite{KS-V}:

\begin{equation}
\label{Polarization}
P^{(\lambda)}=\frac{e}{(2\pi)^{3}}\sum_{n}
 \int_{BZ}d{\bf{k}}
<u_{n\bf{k}}^{(\lambda)}|i\nabla_{\bf{k}}|u_{n\bf{k}}^{(\lambda)}>
\end{equation}

\noindent where $e>0$ is the absolute value of the charge of the electron, $BZ$
is the Brillouin zone of the crystal and $n$ is a discrete index, running over occupied bands.

\noindent Variations of this  dipole per unit volume have been interpreted as
the $\it{integrated}$ electronic $\it{polarisation}$ current \cite{Resta2}.
According to the modern point of view, only variations of polarization are physical observables, $\it{completely~ independent}$ of the periodic charge
distribution. The charge density is represented by the squared modulus  of the wave functions\cite{Resta2}. 
Variations of polarization, instead, have been related to 
a  Berry's phase\cite{KS-V,Resta2} of the system.
We shall try to clarify this statement in what follows.


\section{Geometric phase and V matrix}
\label{CurrentM}

Here it is shown how to develop a calculation scheme in order to model the response of a body, in its
$\it{steady~states}$\cite{primo}, to a static uniform electromagnetic field. 
The model is based on the concepts of group velocity and of conserved geometric phase. 

At this point we point out that we may assume definition
(\ref{equation1}) of being valid even when a uniform static $\it{external}$ 
field, is $\it{present}$ in the system, as it is independent,
i.e. external, to the charge distribution of the quantal body (QB).
In the following we shall always consider the $\it{states}$ of the body of being   
implicitly dependent on external fields.
 
It is already known, in condensed matter physics, that the macroscopic electronic dipole moment per unit volume, of
an insulator is $\it{related}$ to a conserved geometric phase\cite{Resta2,KS-V}.
Here, we show how to obtain a conserved geometric phase making use of a
matrix that we shall call $\bf{V}$ matrix. 
${\bf{V}}$ is obtainable from the expectation values of the group velocity operator\cite{primo}. 

We start our discussion presenting in this section part of the work of Zak\cite{ZakBerry,Zak9}, in order to introduce the reasoning which led us to 
dervive results reported in the following section. Firstly, we derive the conserved geometric phase in absence of external electromagnetic fields 
in the system, being the system an infinite crystal. Let us start writing the Schr$\ddot{o}$dinger equation in the $k-q$ 
representation\cite{Zak,ZakBerry} as follows

\begin{equation}
\label{H-nofield}
\begin{split}
[\frac{1}{2m}(-i\hbar \nabla_{{\bf{q}}}+{\hbar \bf{k}})^{2} + V({\bf{q}})
]u_{n,{\bf{k}}}({\bf{q}})&= H_{0,{\bf{k}}}u_{n,{\bf{k}}}({\bf{q}}) \\&= \epsilon_{n}
u_{n,{\bf{k}}}({\bf{q}})\\ 
\end{split}
\end{equation} 

In analogy to \cite{Berry,Zak9}, if we now derive eq.(\ref{H-nofield}) with respect to $\bf{k}$ we obtain

\begin{equation}
\label{H-nofieldder}
\begin{split}
\frac{\hbar}{m}(-i\hbar \nabla_{{\bf{q}}}+{\hbar \bf{k}})
u_{n,{\bf{k}}}({\bf{q}})+H_{0,{\bf{k}}}\nabla_{{\bf{k}}}u_{n,{\bf{k}}}({\bf{q}}) &=
\nabla_{{\bf{k}}}\epsilon_{n} u_{n,{\bf{k}}}({\bf{q}})
\\&+ \epsilon_{n} \nabla_{{\bf{k}}} u_{n,{\bf{k}}}({\bf{q}})
\end{split}
\end{equation} 

Multiplying eq.(\ref{H-nofieldder}) on the left by $u^{*}_{m}$, dividing by $\hbar$ and integrating with respect to
${\bf{q}}$ in the super-cell we obtain

\begin{equation}
\label{jmn}
{\bf{v}_{mn}}=<u_{m,{\bf{k}}}|{\bf{v}}|u_{n,{\bf{k}}}>
=-i {\bf{} {\bf{} \omega_{mn}}}{\bf{d}}_{mn} + \frac{1}{\hbar} \nabla_{{\bf{k}}}\epsilon_{n} {\bf{} \delta_{mn}}
\end{equation} 

where we write the transition frequencies as ${\bf{} \omega_{mn}}=\frac{\epsilon_{m}-\epsilon_{n}}{\hbar}$ and

\begin{equation}
\label{Vk}
\begin{split}
{\bf{v}}=\frac{(-i\hbar \nabla_{{\bf{q}}}+{\hbar \bf{k}})}{m}
\end{split}
\end{equation} 

is the group velocity operator as shown in ref.{\cite{primo}}.

Coefficients ${\bf{v}_{mn}}$ are the matrix elements of the ${\bf{V}}$ matrix.

In the following, we shall express
$<u_{m,{\bf{k}}}|{\bf{v}}|u_{n,{\bf{k}}}>$ as also  
$<m|{\bf{v}}|n>$ in order to simplify the notation.
 
Matrix elements ${\bf{d}_{mn}}$ are defined as

\begin{equation}
\label{Pmn0}
{\bf{d}_{mn}}=<u_{m,{\bf{k}}}| i\nabla_{{\bf{k}}}|u_{n,{\bf{k}}}>  
\end{equation} 

and because of the normalization condition of the complex wave functions $u_{nk}$ follows
that

\begin{equation}
\label{Pmn1}
{\bf{d}^{*}_{mn}}= -{\bf{d}_{nm}} 
\end{equation} 

also implying that the diagonal fields are real. 
From eq.(\ref{jmn}) follows

\begin{equation}
\label{jmn2}
\begin{split}
{\bf{d}_{mn}} &= i\frac{{\bf{v}_{mn}}}{{\bf{} \omega_{mn}}} ~~  m \neq n \\
{\bf{v}_{nn}} &= \frac{1}{\hbar} \nabla_{{\bf{k}}}\epsilon_{n} ~~  m =  n \\
\end{split}
\end{equation} 

Let us first note that the trace of the velocity operator, evaluated with respect to the occupied bands, is 

\begin{equation}
\label{Trjmn}
{\bf{}Tr}({\bf{v}_{mn}})=\sum_{n} {\bf{v}_{nn}} = \sum_{n} \frac{1}{\hbar}\nabla_{{\bf{k}}}\epsilon_{n}={\bf{v}}({\bf{k}})
\end{equation} 

that is an invariant quantity because of its geometric nature. 
It represents the electronic current density of the
system, in $\bf{k}$ space,  once multiplied by the electronic charge $e$,

\begin{equation}
\label{corrente}
{\bf{j}}({\bf{k}})=e {\bf{}Tr}({\bf{v}_{mn}})=e{\bf{v}}({\bf{k}})
\end{equation} 

By relation (\ref{curl-dnn}) in the appendix, and taking the curl of vectors ${\bf{d}_{nn}}$ it si possible to define 

\begin{equation}
\label{rotPmn}
\nabla_{\bf{k}} \times {\bf{d}_{nn}}= i\int \nabla_{{\bf{k}}}u^{*}_{n,{{\bf{k}}}}({\bf{q}}) \times \nabla_{{\bf{k}}}u_{n,{{\bf{k}}}}({\bf{q}})d{\bf{q}}
\end{equation} 

as already shown by Zak\cite{Zak9}. Also, by equation (\ref{Pmn0}) follows that

\begin{equation}
\label{nablak}
 i\nabla_{\bf{k}}u_{n,{\bf{k}}}({\bf{q}})=\sum_{m}{\bf{d}_{mn}}u_{m,{\bf{k}}}({\bf{q}})  
\end{equation} 

and combined with eq.(\ref{rotPmn}) gives   

\begin{equation}
\label{rotPmn2}
\nabla_{\bf{k}} \times {\bf{d}_{nn}} = i \sum_{m \neq n} {\bf{d}_{mn}} \times {\bf{d}_{nm}} 
\end{equation}

Making use of eq.(\ref{jmn2}) we obtain

\begin{equation}
\label{rotPmn22}
\nabla_{\bf{k}} \times {\bf{d}_{nn}} = i\sum_{m \neq n} {\bf{d}_{mn}} \times
{\bf{d}_{nm}} = i \sum_{m \neq n} \frac{{\bf{v}_{mn}} \times {\bf{v}_{nm}}}{\omega^{2}_{mn}} 
\end{equation} 
 
in agreement with Berry and Zak's derivations \cite{Berry,Zak9}.

We can evaluate then the flux of the field given by eq.(\ref{rotPmn2}), through
the closed surface $S$ of the Brillouin zone in $\bf{k}$ space, containing a closed
curve $C$.

We may either write

\begin{equation}
\label{flusso}
\begin{split}
\int_{S} d{\bf{S}} \cdot {\bf{Im}} \sum_{m \neq n} [\frac{{\bf{v}_{mn}} \times {\bf{v}_{nm}}}{\omega^{2}_{mn}}] 
&=\int_{S} d{\bf{S}} \cdot \nabla_{\bf{k}} \times {\bf{d}_{nn}}\\
&=\oint_{C} {\bf{d}_{nn}} \cdot {d\bf{l}} =\phi_{n}(C) \\
\end{split}
\end{equation}

or in a different way,

\begin{equation}
\label{flusso1}
\begin{split}
\int_{S} d{\bf{S}} \cdot {\bf{Im}}   \sum_{m \neq n} {\bf{d}_{mn}} \times {\bf{d}_{nm}} 
&=\int_{S} d{\bf{S}} \cdot \nabla_{\bf{k}} \times {\bf{d}_{nn}}\\
&=\oint_{C} {\bf{d}_{nn}} \cdot {d\bf{l}} =\phi_{n}(C) \\
\end{split}
\end{equation}

where $\bf{Im}$ stands for imaginary. After a summation over n,  we find

\begin{equation}
\label{sum-flusso}
\sum_{n} \phi_{n}(C) = \Phi(C)
\end{equation}

being $\Phi(C)$ is the conserved geometric phase. 
In view of eq.(\ref{rotPmn22}) we note that $\nabla_{\bf{k}} \times {\bf{d}_{nn}} $ is the
trace of the Berry's curvature\cite{Resta2}, while instead ${\bf{d}_{nn}}$ is Berry's connection\cite{Resta2}.
Let us calculate the averaged trace of fields ${\bf{d}_{nn}}$, integrated over the Brillouin zone in
reciprocal space, or in other words

\begin{equation}
\label{DefP}
\begin{split}
\bar{{\bf{P}}}& = e {\bar{\bf{d}}} \\
&=\frac{e}{(2 \pi)^{3}} \sum^{occ}_{n} \int_{BZ} dk <u_{n{\bf{k}}}| i\nabla_{{\bf{k}}}|u_{n{\bf{k}}}>\\
&=\frac{e}{(2 \pi)^{3}} \int_{BZ} dk {\bf{}Tr}({\bf{d}_{nn}})\\ 
&=e \int_{BZ} \frac{dk}{(2 \pi)^{3}} {\bf{d}} \\ 
\end{split}
\end{equation}

where $e$ is the electronic charge. The circuit integral of the field ${\bf{P}}$ defined as

\begin{equation}
\label{DefP1}
\begin{split}
{\bf{P}}& = e {{\bf{d}}} 
\end{split}
\end{equation}

is   

\begin{equation}
\label{integralechiuso}
\begin{split}
\oint_{C} {\bf{P}} \cdot {d\bf{l}} =e\Phi(C) \\
\end{split}
\end{equation}
 
Eq.(\ref{DefP}) is coincident with results found in
ref.(\cite{KS-V,Resta4})  (see also
\cite{Umari-Pasqua,Umari-Pasqua2,Vanderbilt2,Martin} for its use in numerical calculations).
Because of that we  can associate to $\bar{\bf{P}}$ the meaning of macroscopic dipole moment per unit volume of the
system, in its steady states. 

Berry's phase instead is given by eq.(\ref{integralechiuso}), so that we
may think of a classification of materials, as a function of values of 
$\Phi(C)$, directly related to definition (\ref{DefP1}).

Moreover, we may calculate the conserved phase
$\Phi(C)$, not only by the knowledge of matrix elements of
$V$, as in eq.(\ref{flusso}), but also by the knowledge of matrix elements
${\bf{d}_{mn}}$ by eq.(\ref{flusso1}). 
A discussion about degeneracies of the energy spectrum can be found in references \cite{Berry,Aronov,ZakBerry,Zak9,Thouless}. 

For what stated in \cite{Resta1}, and by 
results given in \cite{primo} the reasoning may be directly
extended to a many-body theory. 

In that context the expression (\ref{Vk}) will be generalized by eq.(18) of \cite{primo}. 

\begin{equation}
\label{Vgen}
{\bf{v}_{mn}}= <m|\frac{1}{\hbar}\nabla_{\bf{k}}H|n>
\end{equation}

being $H$ the $\bf{k}$ dependent Hamiltonian of the problem.

In the following, we extend the problem, to the case of an 
external uniform magnetostatic field present in the system.  

\section{V matrix in magnetic fields}
\label{CurrentMB0}

In this section we generalize the problem to the case of an external uniform magnetostatic field $B^{0}$ present in the system.
The Hamiltonian operator is given by

\begin{equation}
\label{vel-kq6}
\begin{split}
H_{{\bf{k}}} \equiv &[\frac{1}{2m}(-i\hbar \nabla_{{\bf{q}}}+{\hbar \bf{k}}+\frac{e}{c}
  B^{0} \times i\nabla_{{\bf{k}}})^{2} \\ &+ V({\bf{q}})]
\end{split}
\end{equation} 

The corresponding expression of the group velocity operator is then,

\begin{equation}
\label{velocita}
\begin{split}
{\bf{v}}^{(B^{0})}=\frac{1}{\hbar} \nabla_{{\bf{k}}} H_{{\bf{k}}}
\end{split}
\end{equation} 

so that matrix elements of the velocity operator are expressible as follows 
 
\begin{equation}
\label{jmn-B0}
\begin{split}
{\bf{v}^{(B^{0})}_{mn}}=<m|{\bf{v}}^{(B^{0})}|n>&=-i {\bf{} \omega_{mn}} {\bf{d}_{mn}} 
\\&+ \frac{1}{\hbar} \nabla_{{\bf{k}}}\epsilon_{n} {\bf{} \delta_{mn}} 
\\&+ \frac{e}{mc} B^{0} \times {\bf{d}_{mn}}\\
\end{split}
\end{equation}

In analogy with (\ref{corrente}), the current density is

\begin{equation}
\label{correnteB^{0}}
{\bf{j}}({\bf{k}})=e {\bf{}Tr}({\bf{v}_{mn}})=e{\bf{v}}({\bf{k}})+\frac{1}{c} \omega_{cycl} {\bf{b}} \times {\bf{P}}
\end{equation} 

being ${\bf{b}}$ a unit vector parallel to the direction of the magnetic field
and $\omega_{cycl}=\frac{e|B^{0}|}{m}$ the cyclotron frequency of an electron.  
Moreover, by equation (\ref{correnteB^{0}}) and (\ref{Divaxb}) we can calculate the divergence of the current as follows

\begin{equation}
\label{divcorrenteB^{0}}
\nabla \cdot {\bf{j}}({\bf{k}})=e \nabla \cdot {\bf{v}} - \frac{1}{c} \omega_{cycl} {\bf{b}} \cdot \bf{rot} \bf{P}
\end{equation} 

In the following section we shall derive another vectorial field, that is directly related
to the curl of the dipolar field (\ref{DefP1}). For the sake of clarity we report in the text only main
results, while details about our calculations are given in the appendix.

\section{${\bf{J}^{S}}$ field}
\label{Vortex}

Let us define

\begin{equation}
\label{vortice}
 {\bf{v}^{S}_{mn}}=\frac{{\bf{v}^{(B^{0})}_{mn}} \times
  {\bf{v}^{(B^{0})}_{nm}}+{\bf{v}^{(B^{0})}_{nm}} \times
  {\bf{v}^{(B^{0})}_{mn}}}{2{\bf{} \omega_{mn}}}
\end{equation}

where $S$ stands for symmetric combination.
After a summation over $m \neq n$ we obtain 

\begin{equation}
\label{vortice1}
\begin{split}
{\bf{v}^{S}_{n}}&=\sum^{occ}_{m \neq n} [\frac{{\bf{v}^{(B^{0})}_{mn}} \times
  {\bf{v}^{(B^{0})}_{nm}}+{\bf{v}^{(B^{0})}_{nm}} \times
  {\bf{v}^{(B^{0})}_{mn}}}{2{\bf{} \omega_{mn}}}]\\
&=\frac{e}{mc} B^{0} \times {\bf{rot}}~ {\bf{d}_{nn}} \\
\end{split}
\end{equation}

being ${\bf{rot}}~{\bf{d}_{nn}}=\nabla_{\bf{k}} \times {\bf{d}_{nn}}$.

By eq.(\ref{DefP}) is valid the following relation

\begin{equation}
\begin{split}
\label{Js-rotP}
\frac{\omega_{cycl}}{c} {\bf{b}} \times {\bf{rot}}~{\bf{P}}
&= {\bf{J}^{S}} 
\end{split}
\end{equation}

where  

\begin{equation}
{\bf{J}^{S}}=e\sum_{n}{\bf{v}^{S}_{n}}=e {\bf{v}^{S}}
\end{equation}

Definition (\ref{Js-rotP}) also implies that

\begin{equation}
\label{DivJS}
\begin{split}
 \int_{V} dV \nabla \cdot {\bf{J}^{S}} = - {\bf{b}} \cdot \int_{S} {d\bf{S}} \times {\bf{rot}~{\bf{P}}}\\
\end{split}
\end{equation} 

We associate the meaning of a current to the field $\bf{J}^{S}$, whose source
is related to the rotational properties of ${\bf{P}}$. An alternative derivation of the above results can be obtained by the definition of
another vectorial field. By the following relation

\begin{equation}
\label{V1}
\begin{split}
[{\bf{v}^{(B^{0})}_{mn}} \times {\bf{d}_{nm}}] &= -i{\bf{} \omega_{mn}} ({\bf{d}_{mn}} \times
{\bf{d}_{nm}}) \\&+ \frac{e}{mc} (B^{0} \times {\bf{d}_{mn}}) \times {\bf{d}_{nm}}
\end{split}
\end{equation} 

we find

\begin{equation}
\label{PxJ1}
\begin{split}
{\bf{l}^{A}}_{n}&=i\sum_{m \neq n }[{\bf{d}_{nm}} \times {\bf{v}^{(B^{0})}_{mn}}] -
   [{\bf{d}_{mn}} \times {\bf{v}^{(B^{0})}_{nm}}]  
\\&= \frac{e}{mc}B^{0} \times {\bf{rot}~{\bf{d}}_{nn}}\\
\end{split}
\end{equation}

directly derivable by $\bf{d}_{mn}$ and $\bf{v}_{mn}$ matrix elements, it is clear
that $\bf{l}^{A}_{n} \equiv V^{S}_{n}$ and we can define  

\begin{equation}
{\bf{J}^{S}}=e\sum_{n}{\bf{l}^{A}_{n}} 
\end{equation}
.

\section{Appearance of the geometric phase}
Taking the vectorial product of $B^{0}$ and the field ${\bf{v}^{S}_{n}}$ we obtain 
 
\begin{equation}
\label{BxVnn}
\begin{split}
 B^{0} \times  {\bf{v}^{S}_{n}} &=\frac{e|B^{0}|^{2}}{mc} {\bf{b}} \times ({\bf{b}} \times
{\bf{rot}} {\bf{d}_{nn}}) \\
&=\frac{e|B^{0}|^{2}}{mc} ({\bf{b}} \cdot {\bf{rot}}~{\bf{d}_{nn}}){\bf{b}}
- \frac{e|B^{0}|^{2}}{mc}  {\bf{rot}}~{\bf{d}_{nn}}\\
\end{split}
\end{equation}

Let us exclude the cases ${\bf{rot~P}}=0$, and ${\bf{rot~P}}$ parallel to
$\bf{b}$. 

Dividing eq.(\ref{BxVnn}) by $|B^{0}|$, summing over $n$ and calculating the flux of
${\bf{v}^{S}_{n}}$ over the surface $S$ of the Brillouin zone,
containing a closed curve $C$ we obtain

\begin{equation}
\label{BerryfluxBxVnn}
\begin{split}
-[\int_{S} {d\bf{S}} \cdot ({\bf{b}} \times  {\bf{v}^{S}}) ]
&=\frac{\omega_{cycl}}{c}  \int_{S} {d\bf{S}} \cdot {\bf{rot}}~{\bf{d}} 
\\&- \frac{\omega_{cycl}}{c} \int_{S}({\bf{b}} \cdot {\bf{rot}}~{\bf{d}}){\bf{b}} \cdot{d\bf{S}} \\
&=\frac{\omega_{cycl}}{c}\Phi(C)
\\&-\frac{\omega_{cycl}}{c}  \int_{S}({\bf{b}} \cdot{\bf{rot}}~{\bf{d}}){\bf{b}}\cdot{d\bf{S}}\\
\end{split}
\end{equation}

Multiplying eq.(\ref{BerryfluxBxVnn}) by the electronic charge and bearing in mind eq.(\ref{integralechiuso}) we obtain   

\begin{equation}
\label{BerryfluxBxVnnPol}
\begin{split}
{\bf{b}} \cdot \int_{V}dV {\bf{rot}}~ {\bf{J}^{S}} &= -[\int_{S} {d\bf{S}} \cdot ({\bf{b}} \times {\bf{J}^{S}})]
\\&= \frac{\omega_{cycl}}{c}e\Phi(C)
\\&- \frac{\omega_{cycl}}{c} \int_{S}({\bf{b}} \cdot{\bf{rot}}~{\bf{P}}){\bf{b}}\cdot{d\bf{S}}\\
\end{split}
\end{equation}

\section{Low powers of $\frac{1}{c}$}
\label{lowenerg}

Bearing in mind definition (\ref{jmn-B0}), we can write at first order of powers of $\frac{1}{c}$  

\begin{equation}
\label{Vorticello1}
\begin{split}
[{\bf{v}^{(B^{0})}_{mn}} \times {\bf{v}^{(B^{0})}_{nm}}] &\sim \omega^{2}_{mn} {\bf{d}_{mn}} \times {\bf{d}_{nm}}
\\&+i{\bf{} \omega_{mn}}\frac{e}{mc} B^{0} \times ({\bf{d}_{mn}} \times {\bf{d}_{nm}})
\end{split}
\end{equation}

Defining

\begin{equation}
\label{Vorticellantropo}
\begin{split}
{\bf{V}^{A}_{mn}}=[\frac{{\bf{v}^{(B^{0})}_{mn}} \times
    {\bf{v}^{(B^{0})}_{nm}}-{\bf{v}^{(B^{0})}_{nm}} \times
    {\bf{v}^{(B^{0})}_{mn}}}{2}] 
\end{split}
\end{equation}

dividing by $\omega^{2}_{mn}$,  and summing over $m \neq n$ we obtain 

\begin{equation}
\label{Vorticello2}
\begin{split}
{\bf{V}^{A}_{n}}&=i\sum^{occ}_{m \neq n}[\frac{{\bf{v}^{(B^{0})}_{mn}} \times
    {\bf{v}^{(B^{0})}_{nm}}-{\bf{v}^{(B^{0})}_{nm}} \times
    {\bf{v}^{(B^{0})}_{mn}}}{2\omega^{2}_{mn}}] \\&= {\bf{rot}}~{\bf{d}_{nn}}
\end{split}
\end{equation}

Evaluating the flux over the surface $S$ of the Brillouin zone we find 

\begin{equation}
\label{Vorticello3}
\begin{split}
\int_{S}d{\bf{S}} \cdot [{\bf{V}^{A}_{n}}] = \Phi_{n}(C)
\end{split}
\end{equation}

If we instead consider also terms containing second powers of $\frac{1}{c}$, we
instead obtain 

\begin{equation}
\label{Vorticello4}
\begin{split}
\int_{S}d{\bf{S}} \cdot [{\bf{V}^{A}_{n}}] &= \Phi_{n}(C) \\
&+ \frac{1}{c^{2}}\sum_{m \neq n } \int_{S} d{\bf{S}} \cdot
[\frac{\omega_{cycl}^{2}}{\omega^{2}_{mn}} (\bf{b} \cdot ({\bf{d}_{mn}} \times {\bf{d}_{nm}}))\bf{b}]
\end{split}
\end{equation}
 
Equations (\ref{flusso}) and (\ref{BerryfluxBxVnnPol}), in conjunction with eq.(\ref{integralechiuso}) and eq.(\ref{Js-rotP}), clearly  
show that if the vectorial field ${\bf{d}}$ is conservative (irrotational, $\bf{rot}~{\bf{d}} =\bf{0}$), then the system would
acquire a vanishing Berry's phase, and the vectorial field ${\bf{J}^{S}}$ would become vanishing by eq.(\ref{Js-rotP}). 
Moreover, if ${\bf{d}}$ is conservative, the divergence of the current ${\bf{j}(k)}$ will not be affected by the latter. 
In the case of ${\bf{d}}$ not being conservative the flux of ${\bf{j}(k)}$ over the surface of the Brillouin zone will be 
instead proportional to the projection, over the direction of the applied magnetostatic field, of the circulation of  ${\bf{d}}$ (or equivalently 
${\bf{P}}$) over the surface of the Brillouin zone.

\section{Power in a uniform static electromagnetic field}
\label{potenza}

The group velocity is not explicitly dependent on a uniform
electrostatic field $E_{0}$ eventually present in the system, so that all the results obtained above are equally right even if
$E_{0}$ is present in the system. 
Bearing in mind eq.(\ref{correnteB^{0}}) and eq.(\ref{DefP}), we can express the work per unit
time and per unit volume done by the electromagnetic field $(E_{0},B^{0})$ on the body as 

\begin{equation}
\label{power} 
\begin{split}
W&=E_{0} \cdot J  =  E^{0} \cdot \int_{V} dV  {\bf{j}}({\bf{k}}) \\
& =e E^{0} \cdot \int_{V} dV  {\bf{v}}({\bf{k}}) +  \frac{4 \pi e^{2}}{mc^{2}}
{\bf{S}} \cdot \int_{V} dV {\bf{d}} \\
&=e E^{0} \cdot \bar{{\bf{v}}} +  \frac{4 \pi e^{2}}{mc^{2}}
{\bf{S}} \cdot \bar{{\bf{d}}} \\
\end{split}
\end{equation}

being $\bf{S}=\frac{c}{4 \pi} (E_{0}\times B^{0})$ the Poynting
vector\cite{Jackson,Pauli} associated to the electromagnetic field and $V$ the volume of the Brillouin zone in $k$-space.

\section{Conclusions}
\label{conclusioni}

The reported non-relativistic calculation scheme, may be useful for practical calculations
of macroscopic electromagnetic properties of a quantal body interacting with
a static external uniform electromagnetic field. 
It has been shown how to calculate the power density, the macroscopic polarization ${\bar{\bf{P}}}$, and the electronic 
current induced by the presence of a uniform magnetostatic field $B^{0}$, making use of the concept of group velocity. 
It has been shown the reasoning in the case of the one-body approximation, and by construction it would apply to every 
first-principles calculation scheme\cite{Resta2,KS-V,Resta5,Resta-ferro}, based on
periodic boundary conditions within a DFT scheme\cite{Sham}. 
The above discussion can be extended to a 'many-body' theory because of what stated in\cite{Resta1}.
Eq.(\ref{flusso}) may help us for a theoretical classification of crystalline materials via the values of $\Phi(C)$, that are directly dependent on 
the dielectric properties of the  material under consideration via eq.(\ref{integralechiuso}). 
We may also think that dielectric and magnetic properties of the crystalline material determine its geometrical properties too. 
In fact eq.(\ref{integralechiuso}),eq.(\ref{Js-rotP}) show that if the vectorial field ${\bf{d}}$ is conservative (irrotational, 
$\bf{rot}~{\bf{d}} =\bf{0}$), then the system would acquire a vanishing Berry's phase and the vectorial field ${\bf{J}^{S}}$ would become 
vanishing by eq.(\ref{Js-rotP}). 
Nevertheless a proper theoretical solution of the problem requires a theoretical study of the ionic response as well as the electronic one, 
that we leave to a future work.     
The model presented above allows us to find several useful physical quantities in a clear and simple
way, amenable of a straightforward physical interpretation. 

\section{Appendix} 
\subsection{Vector~Identities}
\label{VI}

\begin{equation}
\label{curl-dnn}
\nabla \times (f{\bf{A}}) 
= f \nabla \times {\bf{A}} +  (\nabla f) \times {\bf{A}} 
\end{equation}

\begin{equation}
\label{rel1}
{\bf{A}} \times ({\bf{B}} \times {\bf{C}}) = ({\bf{A}} \cdot {\bf{C}}){\bf{B}}
- ({\bf{A}} \cdot {\bf{B}}){\bf{C}}
\end{equation}

 \begin{equation}
\label{rel2}
({\bf{A}}\times {\bf{B}} ) \times ({\bf{C}} \times {\bf{D}}) = (({\bf{A}}
\times {\bf{B}}) \cdot {\bf{D}}){\bf{C}} 
- (({\bf{A}} \times {\bf{B}}) \cdot {\bf{C}}){\bf{D}} 
\end{equation}

\begin{equation}
\label{Divaxb}
(\nabla \cdot {\bf{A}}\times {\bf{B}} ) =  {\bf{B}} \cdot \nabla \times
      {\bf{A}} - {\bf{A}} \cdot \nabla \times  {\bf{B}} 
\end{equation}

\section{$J^{S}$ field}
\label{VSA}

Let us evaluate the following quantity 

\begin{equation}
\label{expr1AA}
\begin{split}
{\bf{v}^{(B^{0})}_{mn}} \times {\bf{v}^{(B^{0})}_{nm}} 
&= \omega^{2}_{mn} {\bf{d}_{mn}} \times {\bf{d}_{nm}} 
\\&+\frac{e^{2}}{m^{2}c^{2}} [(B^{0}\times {\bf{d}_{mn}})\times(B^{0} \times {\bf{d}_{nm}})]
\\&-i{\bf{} \omega_{mn}} [{\bf{d}_{mn}} \times \frac{e}{mc} (B^{0} \times {\bf{d}_{nm}})]
\\& +i{\bf{} \omega_{nm}} [{\bf{d}_{nm}} \times \frac{e}{mc} (B^{0} \times {\bf{d}_{mn}})]\\
\end{split}
\end{equation}

From eq.(\ref{expr1AA}), making use of the vector identity (\ref{rel1}), and the fact that
${\bf{} \omega_{mn}}=-{\bf{} \omega_{nm}}$ we can recast the last two terms of eq.(\ref{expr1AA}) as follows 

\begin{equation}
\label{recast1-A}
\begin{split}
&-i{\bf{} \omega_{mn}} [{\bf{d}_{mn}} \times \frac{e}{mc} (B^{0} \times {\bf{d}_{nm}})]\\
&+i {\bf{} \omega_{nm}}[{\bf{d}_{nm}} \times \frac{e}{mc} (B^{0} \times {\bf{d}_{mn}})]\\
&=i{\bf{} \omega_{mn}}\frac{e}{mc} B^{0} \times ({\bf{d}_{mn}} \times {\bf{d}_{nm}})\\
\end{split}
\end{equation}

so that

\begin{equation}
\label{expr1-2-A}
\begin{split}
{\bf{v}^{(B^{0})}_{mn}} \times {\bf{v}^{(B^{0})}_{nm}} 
&= \omega^{2}_{mn} {\bf{d}_{mn}} \times {\bf{d}_{nm}} \\
&+i{\bf{} \omega_{mn}}\frac{e}{mc} B^{0} \times ({\bf{d}_{mn}} \times {\bf{d}_{nm}})\\
&+\frac{e^{2}}{m^{2}c^{2}} [(B^{0}\times {\bf{d}_{mn}})\times(B^{0} \times {\bf{d}_{nm}})]\\
\end{split}
\end{equation}

Making use of the vector identity (\ref{rel2}) and 

\begin{equation}
\label{vecI3-Aprop}
({\bf{A}} \times {\bf{B}}) \cdot {\bf{A}} = 0
\end{equation}
 
we can recast eq.(\ref{expr1-2-A}), as follows

\begin{equation}
\label{exprs1-3-A}
\begin{split}
{\bf{v}^{(B^{0})}_{mn}} \times {\bf{v}^{(B^{0})}_{nm}} 
&=\omega^{2}_{mn} {\bf{d}_{mn}} \times {\bf{d}_{nm}}
\\&+\frac{e^{2}}{m^{2}c^{2}} [ (B^{0}\times {\bf{d}_{mn}}) \times (B^{0}
  \times  {\bf{d}_{nm}}) ]
\\&+i{\bf{} \omega_{mn}}\frac{e}{mc} B^{0} \times ({\bf{d}_{mn}} \times {\bf{d}_{nm}})
\\&=\omega^{2}_{mn} {\bf{d}_{mn}} \times {\bf{d}_{nm}}
\\&+\frac{e^{2}}{m^{2}c^{2}} [((B^{0}\times {\bf{d}_{mn}}) \cdot
  {\bf{d}_{nm}})B^{0} 
\\&- ((B^{0} \times {\bf{d}_{mn}}) \cdot  B^{0}) {\bf{d}_{nm}}   ]\\
&+i{\bf{} \omega_{mn}}\frac{e}{mc} B^{0} \times ({\bf{d}_{mn}} \times {\bf{d}_{nm}})\\
&=\omega^{2}_{mn} {\bf{d}_{mn}} \times {\bf{d}_{nm}}
\\&+\frac{e^{2}}{m^{2}c^{2}} [((B^{0}\times {\bf{d}_{mn}}) \cdot {\bf{d}_{nm}})B^{0}]\\
&+i{\bf{} \omega_{mn}}\frac{e}{mc} B^{0} \times ({\bf{d}_{mn}} \times {\bf{d}_{nm}})\\
&=\omega^{2}_{mn} {\bf{d}_{mn}} \times {\bf{d}_{nm}}
\\&+\frac{e^{2}}{m^{2}c^{2}} (B^{0}\cdot ({\bf{d}_{mn}} \times {\bf{d}_{nm}}))B^{0}\\
&+i{\bf{} \omega_{mn}}\frac{e}{mc} B^{0} \times ({\bf{d}_{mn}} \times {\bf{d}_{nm}})\\
\end{split}
\end{equation}

and define the following quantity,

\begin{equation}
\label{VortexA}
\begin{split}
i\sum_{m \neq n} \frac{{\bf{v}^{(B^{0})}_{mn}} \times {\bf{v}^{(B^{0})}_{nm}}}{\omega^{2}_{mn}} 
&={{\bf{rot}}~{\bf{d}_{nn}}}
\\&-\sum_{m \neq n} \frac{\omega_{cycl}}{{\bf{} \omega_{mn}}c} 
{\bf{b}} \times ({\bf{d}_{mn}} \times {\bf{d}_{nm}})
\\ &+i\sum_{m \neq n} \frac{{\omega^{2}_{cycl}}}{c^{2} \omega^{2}_{mn}} 
({\bf{b}} \cdot ({\bf{d}_{mn}} \times {\bf{d}_{nm}})){\bf{b}}\\
\end{split}
\end{equation}

being $\bf{b}$ a unit vector in the direction of the magnetic axis and
$\omega_{cycl}$ is the electronic cyclotron frequency per unit magnetic field,
defined as $\frac{\omega_{cycl}}{|B^{0}|}=\frac{e}{m}$ .

We can write

\begin{equation}
\label{VortexAA}
\begin{split}
[{\bf{v}^{(B^{0})}_{mn}} \times {\bf{v}^{(B^{0})}_{nm}}] &=\omega^{2}_{mn} {\bf{d}_{mn}} \times {\bf{d}_{nm}}
\\&+i{\bf{} \omega_{mn}}\frac{e}{mc} B^{0} \times ({\bf{d}_{mn}} \times {\bf{d}_{nm}})
\\&+\frac{e^{2}}{m^{2}c^{2}} (B^{0}\cdot ({\bf{d}_{mn}} \times {\bf{d}_{nm}}))B^{0}\\
\end{split}
\end{equation}

and

\begin{equation}
\label{Vortex1AA}
\begin{split}
[{\bf{v}^{(B^{0})}_{nm}} \times {\bf{v}^{(B^{0})}_{mn}}] 
&= \omega^{2}_{nm} {\bf{d}_{nm}} \times {\bf{d}_{mn}}
\\&+i{\bf{} \omega_{nm}}\frac{e}{mc} B^{0} \times ({\bf{d}_{nm}} \times {\bf{d}_{mn}})
\\&+\frac{e^{2}}{m^{2}c^{2}} (B^{0}\cdot ({\bf{d}_{nm}} \times {\bf{d}_{mn}}))B^{0}
\\&=-\omega^{2}_{nm} {\bf{d}_{mn}} \times {\bf{d}_{nm}}
\\&+i{\bf{} \omega_{mn}}\frac{e}{mc} B^{0} \times ({\bf{d}_{mn}} \times {\bf{d}_{nm}})
\\&-\frac{e^{2}}{m^{2}c^{2}} (B^{0}\cdot ({\bf{d}_{mn}} \times {\bf{d}_{nm}}))B^{0}
\\&=-[{\bf{v}^{(B^{0})}_{mn}} \times {\bf{v}^{(B^{0})}_{nm}}] 
\\&+ 2i{\bf{} \omega_{mn}}\frac{e}{mc} B^{0} \times ({\bf{d}_{mn}} \times {\bf{d}_{nm}})\\
\end{split}
\end{equation}

eq.(\ref{VortexAA}) and eq.(\ref{Vortex1AA}) imply then 

\begin{equation}
\label{vorticeAAA}
\begin{split}
 {\bf{v}^{S}_{mn}}&=\frac{{\bf{v}^{(B^{0})}_{mn}} \times
  {\bf{v}^{(B^{0})}_{nm}}+{\bf{v}^{(B^{0})}_{nm}} \times
  {\bf{v}^{(B^{0})}_{mn}}}{2{\bf{} \omega_{mn}}}
\\&=i\frac{e}{mc} B^{0} \times ({\bf{d}_{mn}} \times {\bf{d}_{nm}})
\end{split}
\end{equation}

Also, summing over $m \neq n$ we find 

\begin{equation}
\label{vortice1AAA}
\begin{split}
{\bf{v}^{S}_{n}}&=\sum_{m \neq n} [\frac{{\bf{v}^{(B^{0})}_{mn}} \times
  {\bf{v}^{(B^{0})}_{nm}}+{\bf{v}^{(B^{0})}_{nm}} \times
  {\bf{v}^{(B^{0})}_{mn}}}{2{\bf{} \omega_{mn}}}]\\
&=\frac{e}{mc} B^{0} \times i\sum_{m \neq n}({\bf{d}_{mn}} \times {\bf{d}_{nm}})
=\frac{e}{mc} B^{0} \times {\bf{rot}} {\bf{d}_{nn}} \\
\end{split}
\end{equation}

eq.(\ref{vortice1AAA}) and eq.(\ref{DefP})implies

\begin{equation}
\label{vorticetauAA}
\begin{split}
\frac{\omega_{cycl}}{c} {\bf{b}} \times {\bf{rot}}{\bf{P}}
&= {\bf{J}^{S}} 
\end{split}
\end{equation}

where  

\begin{equation}
{\bf{P}}={e\bf{d}}=e\sum_{n}{\bf{d}_{nn}} 
\end{equation}

and 

\begin{equation}
{\bf{J}^{S}}=e \sum_{n} {\bf{v}^{S}_{n}}= e {\bf{v}^{S}} 
\end{equation}

\subsection{${\bf{l}^{A}}$ field}
\label{tauc}

Let us write the following quantity

\begin{equation}
\label{V1A}
\begin{split}
[{\bf{v}^{(B^{0})}_{mn}} \times {\bf{d}_{nm}}] &= -i{\bf{} \omega_{mn}} ({\bf{d}_{mn}} \times
{\bf{d}_{nm}}) \\&+ \frac{e}{mc} (B^{0} \times {\bf{d}_{mn}}) \times {\bf{d}_{nm}}\\
\end{split}
\end{equation} 

We find 

\begin{equation}
\label{JxP}
    i\sum_{m \neq n }[{\bf{v}^{(B^{0})}_{mn}} \times
    {\bf{d}_{nm}}] -[{\bf{v}^{(B^{0})}_{nm}} \times
    {\bf{d}_{mn}}]  = \frac{e}{mc}B^{0} \times {\bf{rot}~{\bf{d}}_{nn}}
\end{equation}

In fact by making use of definition (\ref{rotPmn2}), the vector identity (\ref{rel1}),  eq.(\ref{Pmn1}),
and the relation  ${\bf{} \omega_{mn}}=-{\bf{} \omega_{nm}}$, we find

\begin{equation}
\label{Vm-VvA}
\begin{split}
    &[{\bf{v}^{(B^{0})}_{mn}} \times {\bf{d}_{nm}}] -[{\bf{v}^{(B^{0})}_{nm}}\times {\bf{d}_{mn}}]=  \\
    & = -i{\bf{} \omega_{mn}} ({\bf{d}_{mn}} \times {\bf{d}_{nm}}) + \frac{e}{mc} (B^{0} \times {\bf{d}_{mn}}) \times {\bf{d}_{nm}}
     \\& -[ -i{\bf{} \omega_{nm}} ({\bf{d}_{nm}} \times {\bf{d}_{mn}}) + \frac{e}{mc}(B^{0} \times {\bf{d}_{nm}}) \times {\bf{d}_{mn}}]\\
    &= -i{\bf{} \omega_{mn}} ({\bf{d}_{mn}} \times {\bf{d}_{nm}}) + \frac{e}{mc} (B^{0} \times {\bf{d}_{mn}}) \times {\bf{d}_{nm}}
     \\& -[ i{\bf{} \omega_{mn}} ({\bf{d}_{nm}} \times {\bf{d}_{mn}}) + \frac{e}{mc}(B^{0} \times {\bf{d}_{nm}}) \times {\bf{d}_{mn}}]\\
    &= -i{\bf{} \omega_{mn}} ({\bf{d}_{mn}} \times {\bf{d}_{nm}}) + \frac{e}{mc} (B^{0} \times {\bf{d}_{mn}}) \times {\bf{d}_{nm}}
    \\&  -[-i{\bf{} \omega_{mn}} ({\bf{d}_{mn}} \times {\bf{d}_{nm}}) + \frac{e}{mc}(B^{0} \times {\bf{d}_{nm}}) \times {\bf{d}_{mn}}]\\
    &= -i{\bf{} \omega_{mn}} ({\bf{d}_{mn}} \times {\bf{d}_{nm}}) + \frac{e}{mc} (B^{0} \times {\bf{d}_{mn}}) \times {\bf{d}_{nm}}
     \\&  +i{\bf{} \omega_{mn}}  ({\bf{d}_{mn}} \times {\bf{d}_{nm}}) - \frac{e}{mc}(B^{0}  \times {\bf{d}_{nm}}) \times {\bf{d}_{mn}} \\
    &= \frac{e}{mc} (B^{0} \times {\bf{d}_{mn}}) \times {\bf{d}_{nm}}  - \frac{e}{mc}(B^{0} \times {\bf{d}_{nm}}) \times {\bf{d}_{mn}}\\
    &=\frac{e}{mc} [-{\bf{d}_{nm}} \times (B^{0} \times {\bf{d}_{mn}}) + {\bf{d}_{mn}} \times (B^{0} \times {\bf{d}_{nm}})]\\
    &=\frac{e}{mc} [-({\bf{d}_{nm}} \cdot {\bf{d}_{mn}}) B^{0} + ({\bf{d}_{nm}} \cdot B^{0}){\bf{d}_{mn}} 
    \\&                +({\bf{d}_{mn}} \cdot {\bf{d}_{nm}}) B^{0} - ({\bf{d}_{mn}} \cdot B^{0}){\bf{d}_{nm}}]\\
    &=\frac{e}{mc} [({\bf{d}_{nm}} \cdot B^{0}){\bf{d}_{mn}}  - ({\bf{d}_{mn}} \cdot B^{0}){\bf{d}_{nm}}]\\
    &=\frac{e}{mc}  B^{0} \times ({\bf{d}_{mn}} \times {\bf{d}_{nm}})\\
\end{split}
\end{equation}

By definition (\ref{jmn-B0}) ane eq.(\ref{Pmn1}) follows that 

\begin{equation}
\sum_{m \neq n }[{\bf{d}_{nm}} \times {\bf{v}^{(B^{0})}_{mn}}] -   [{\bf{d}_{mn}} \times {\bf{v}^{(B^{0})}_{nm}}] 
\end{equation}
  
is a pure imaginary field.

Also we may re-write eq.(\ref{JxP}) as

\begin{equation}
\label{PxJA}
\begin{split}
{\bf{l}^{A}}&=i\sum_{m \neq n }[{\bf{d}_{nm}} \times {\bf{v}^{(B^{0})}_{mn}}] -
   [{\bf{d}_{mn}} \times {\bf{v}^{(B^{0})}_{nm}}]  
\\&= \frac{e}{mc}B^{0} \times {\bf{rot}~{\bf{d}}_{nn}}
\end{split}
\end{equation}

We can evaluate the following flux over the surface of BZ (Brillouin zone) in
${\bf{k}}$ space, containing a closed curve $C$,

\begin{equation}
\label{ConsA}
\begin{split}
&= \sum_{n} [\int_{S} d{\bf{S}} \cdot  {\bf{l}^{A}}_{n}]
\\&= \sum_{n} \omega_{cycl} \int_{S} d{\bf{S}} \cdot ({\bf{b}}\times {\bf{rot}~{\bf{d}_{nn}}})\\
&= \sum_{n} \omega_{cycl} \int_{V} dV \nabla \cdot ({\bf{b}}\times {\bf{rot}~{\bf{d}_{nn}}})\\
&=- \sum_{n} \omega_{cycl} \int_{V} dv {\bf{b}} \cdot (\nabla \times {\bf{rot}~{\bf{d}_{nn}}})\\
&=- \omega_{cycl} \int_{V} dV {\bf{b}} \cdot (\nabla \times {\bf{rot}~{\bf{d}}})\\
&=- \omega_{cycl} {\bf{b}} \cdot \int_{V} dV (\nabla \times {\bf{rot}~{\bf{d}}})\\
&=- \omega_{cycl} {\bf{b}} \cdot \int_{S} {d\bf{S}} \times {\bf{rot}~{\bf{d}}}\\
\end{split}
\end{equation}

where in the third line has been used Gauss's theorem (divergence theorem), 
and $\omega_{cycl}=\frac{e}{m}$ is the cyclotron frequency per unit magnetic field and 
$B^{0}={|B^{0}|}{\bf{b}}$; 
being ${\bf{b}}$ the direction of the magnetic axis.
Equations (\ref{vorticetauAA}) and (\ref{ConsA}) imply also 

\begin{equation}
\label{DviJSA}
\begin{split}
 \int_{V} dV \nabla \cdot {\bf{J}^{S}} = - {\bf{b}} \cdot \int_{S} {d\bf{S}} \times {\bf{rot}~{\bf{P}}}\\
\end{split}
\end{equation} 

\bibliography{refs}

\end{document}